\documentstyle[epsfig]{aipproc}

\begin{document}
\title{Explosions and Outflows during Galaxy Formation}

\author{Hugo Martel$^*$ and Paul R. Shapiro$^*$}
\address{$^*$Department of Astronomy, University of Texas,
         Austin, TX 78712}

\maketitle

\begin{abstract}
We consider an explosion at the center of a
halo which forms at the intersection of filaments inside
a cosmological pancake, a convenient
test-bed model for galaxy formation. $\rm ASPH/P^3M$ simulations
reveal that such explosions are anisotropic. The energy and metals
are channeled into the low density regions, away from the pancake. 
The pancake remains
essentially undisturbed, even if the explosion is strong enough
to blow away all the gas located inside the 
halo and reheat the
IGM surrounding the pancake. 
Infall quickly replenishes this ejected gas 
and gradually restores the gas fraction as the halo continues
to grow. Estimates of the collapse epoch and  
SN energy-release for galaxies of different
mass in the CDM model can relate these results to scale-dependent
questions of blow-out and blow-away and their implication for 
early IGM heating and metal enrichment 
and the creation of gas-poor dwarf galaxies.
\end{abstract}

\section*{INTRODUCTION}

The release of energy that occurs during galaxy formation 
can have important consequences.
We present 3D gas dynamical simulations of the effect of energy release
by supernovae (SNe) on the
evolution of the halo in which the explosions take place, the
surrounding large-scale structure, and
the IGM. 
 
\begin{figure}[b!] 
\centerline{\epsfig{file=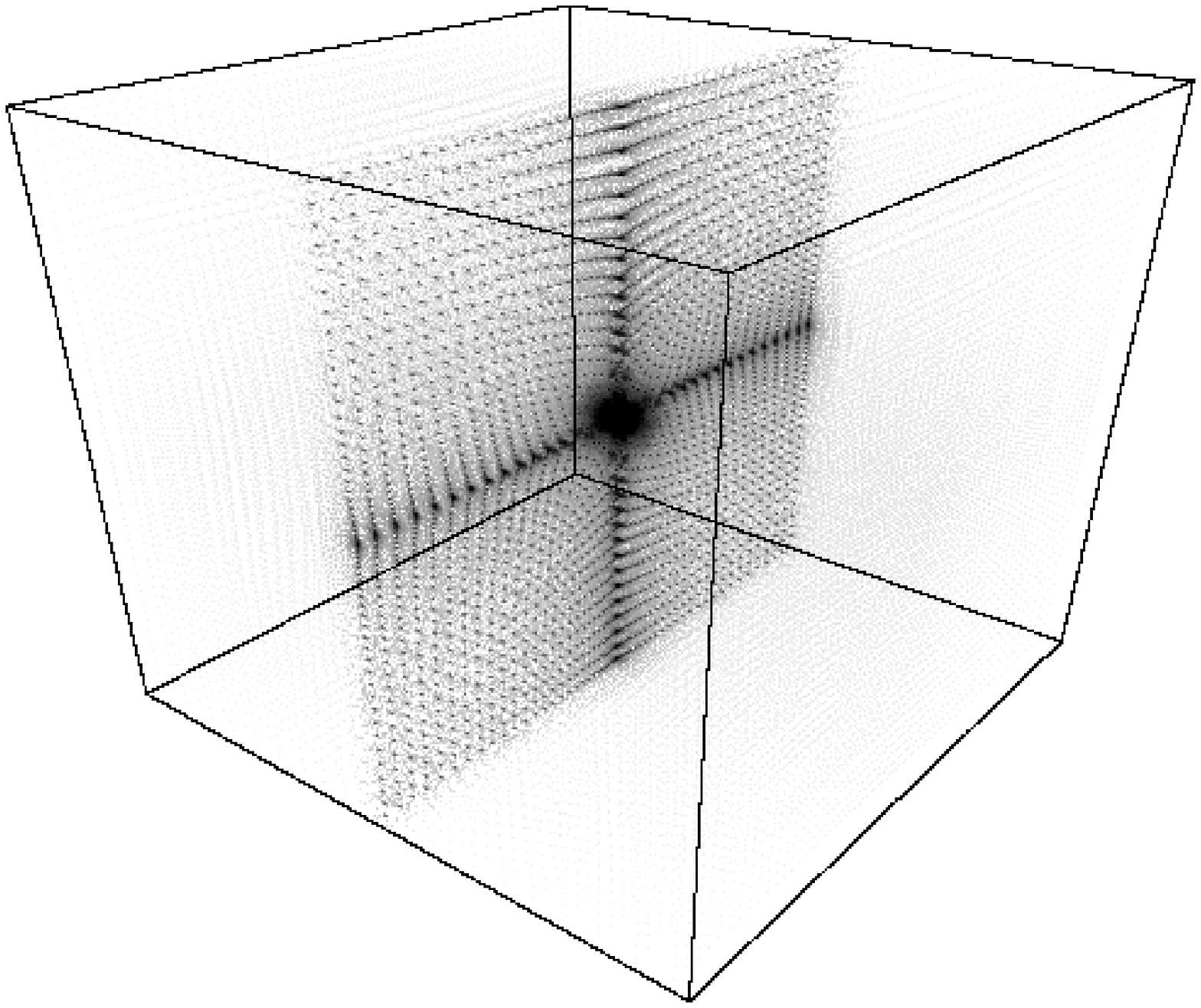,width=2.in}
\epsfig{file=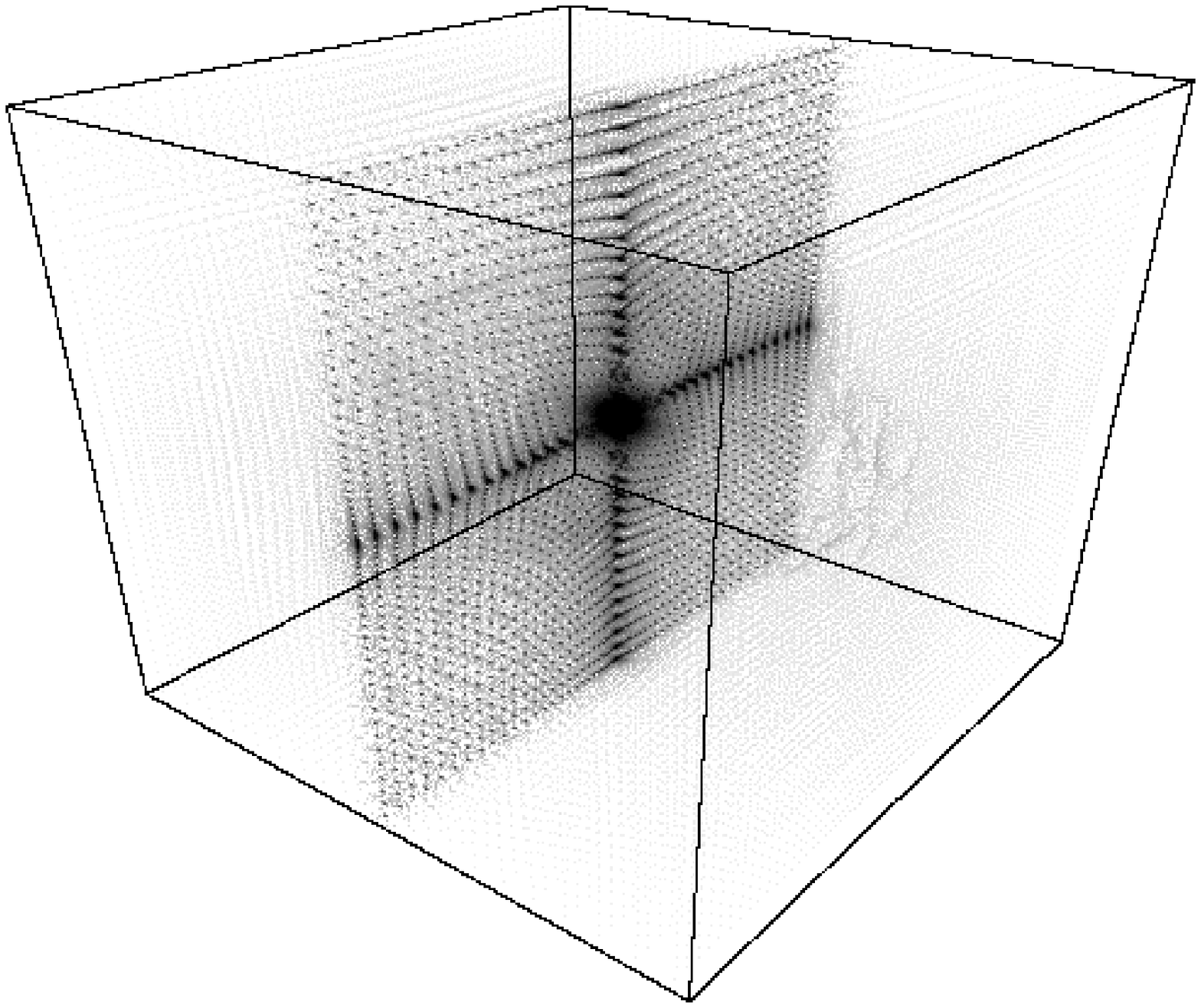,width=2.in}
\epsfig{file=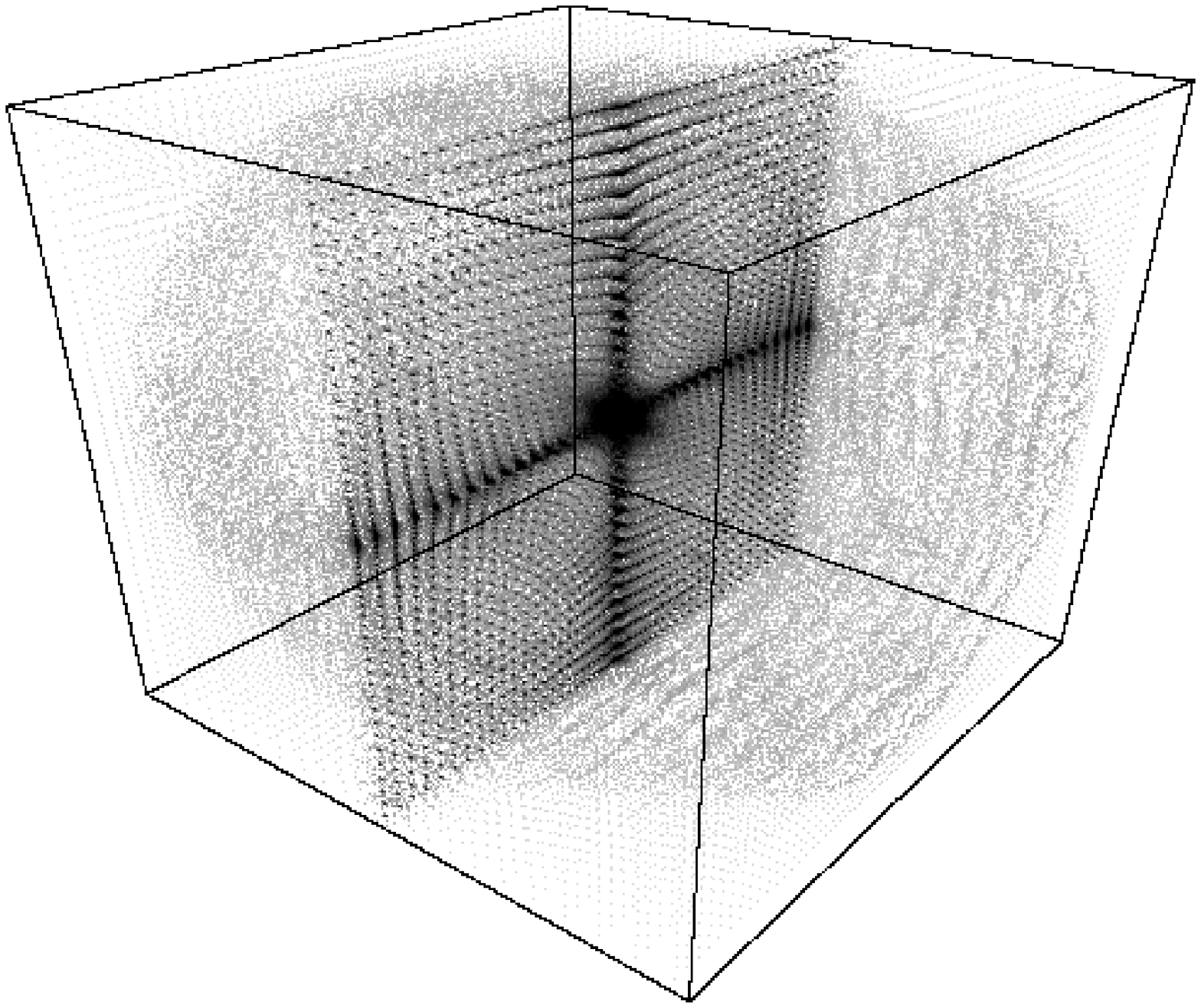,width=2.in}}
\caption{Gas distribution at $a/a_c=3$. (left) no explosion 
case ($\chi=0$);
(middle) explosion case with $\chi=100$; (right) explosion case with 
$\chi=1000$.}
\label{fig1}
\end{figure}

Structure 
formation from Gaussian random noise is highly anisotropic,
favoring pancakes and filaments over quasi-spherical
objects. However, pancake fragmentation
results in the formation of
quasi-spherical halos\cite{alv01,alv02,mar01,val97} with density profiles 
similar to the universal
profile\cite{nfw97} found in CDM simulations.
This suggests that 
pancake fragmentation may be used to study galaxy formation.
This provides a good compromise between simulations of 
structure formation in CDM models, with limited resolution, and
those of isolated virialized objects, which ignore
cosmological initial and boundary conditions.
We use our $\rm ASPH/P^3M$ method to simulate the
formation of a halo at the intersection of two filaments
in the plane of a cosmological pancake which collapses
at scale factor $a=a_c$, in
an $\Omega_0=1$ universe
with baryon fraction $\Omega_{\rm B}=0.03$ 
(see \cite{exp01,exp02} for further description).
These simulations also describe early galaxy formation in a flat,
$\Lambda$CDM model.
The explosion is induced at
scale factor $a_{\rm exp}/a_c\cong2$,
when the central gas density first exceeds
$\rho_{\rm gas}/\langle\rho_{\rm gas}\rangle=1000$,
by boosting the thermal energy of this central gas to
the level 
$E_{\rm exp}=\chi{\cal E}_{\rm halo}$, where ${\cal E}_{\rm halo}$ is the
total thermal energy of gas in the halo, for
four cases: $\chi=0$, 10, 100, and 1000. 


\section*{RESULTS}

\begin{figure}[b!] 
\centerline{\epsfig{file=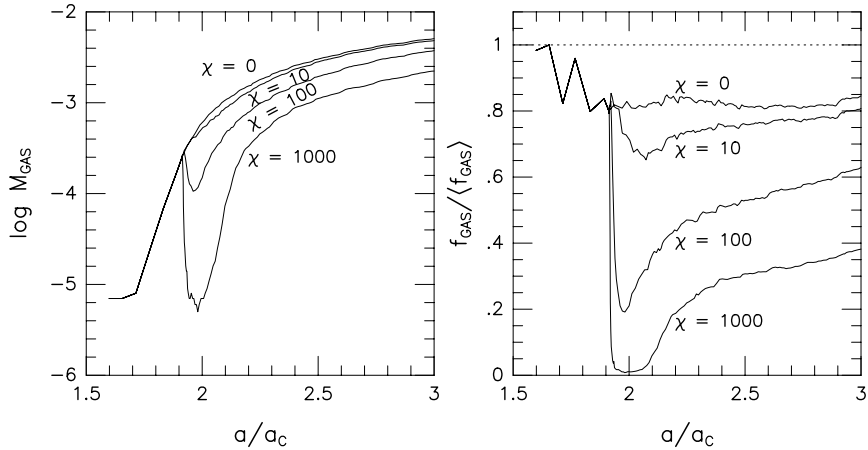,height=2.4in}}
\vspace{10pt}
\caption{(Left) gas mass $M_{\rm gas}$
(in computational units $M_{\rm box}=\bar\rho\lambda_p^3=1$) 
in the central halo (sphere of mean overdensity 200), versus $a/a_c$,
for different explosion intensities, as labelled.
(Right) gas mass fraction $f_{\rm gas}$ inside halo,
divided by the universal gas mass fraction 
$\langle f_{\rm gas}\rangle=\Omega_{\rm B}/\Omega_0$, versus $a/a_c$.}
\label{fig2}
\end{figure}

Figure 1 shows the pancake-filament-halo
structure at $a/a_c=3$, with and without
explosions. For all cases, the dark matter distribution is
essentially the same as the gas distribution for $\chi=0$.
The pancake and filaments ensure that a highly anisotropic
explosion results, channelling the energy and mass ejection
along the symmetry axis.
The ability of explosions to eject metal-enriched gas 
depends on $\chi$. For
$\chi=10$, relatively little gas is blown 
out of the halo, and is quickly re-accreted. 
For $\chi>100$, most
of the halo gas is blown out. However, as shown in Figure~2,
infall along the pancake plane continues despite the explosion and
replenishes the ejected halo gas very
efficiently, although the final gas fraction $f_{\rm gas}$
is below the cosmic mean value by an amount which increases with
increasing $\chi$. Even the most energetic explosions fail to disturb 
the pancake and filaments.

Each dimensionless, scale-free simulation for a given $\chi$
can be applied to
any particular halo mass $M_{\rm halo}$ at any epoch
by adjusting the values of
$\lambda_p$ (the wavelength of the primary pancake) and collapse
redshift $z_c$ when converting to physical units.
We can also use Milky Way (MW) star formation efficiencies and IMF to 
estimate
a typical value $\chi_*(\lambda_p,z_c)$
of the explosion parameter $\chi$, for each $\lambda_p$ and $z_c$.
$\chi\gtrsim100$ is required to eject gas and metals
into the IGM. This matches MW efficiencies
only for $M_{\rm halo}\lesssim10^7M_\odot$, for both
cluster-normalized SCDM (i.e. $\lambda_p<0.2\,\rm Mpc$) and
{\sl COBE}-normalized $\Lambda$CDM ($\Omega_0=0.3$,
$\lambda_0=0.7$, $h=0.7$) (i.e. $\lambda_p<0.3\,\rm Mpc$).
However, gas
replenishment by continued infall may enable
a single halo to contribute multiple outbursts.
Halos $<10^{10}M_\odot$
form early enough to cause heavy element distribution prior to $z=3$.
Only the smallest mass objects, however, can expect to do so
if limited to MW efficiencies. Such small
mass halos are also the ones which are most likely to form early
enough that they may explode {\it before}
the reionization of the IGM is complete, after which
the enhanced IGM pressure inhibits gas ejection.
Otherwise efficiencies greatly exceeding
MW values are required.

A final determination of the success
or failure of heavy element distribution by SN explosions
in low-mass halos forming at high redshift
will depend sensitively on the efficiencies of star formation and
SN energy release, and the relative timing of these
explosions versus universal reionization. 
Our results suggest that
heavy element distribution at the observed level of
$\approx10^{-3}$ solar in the IGM could have been accomplished
most efficiently by low-mass objects
prior to the completion of universal reionization.
(NASA ATP grants NAG5-7363 and
NAG5-7821, NSF grant ASC-9504046, and Texas Advanced Research Program
grant 3658-0624-1999).


\begin{references}
\bibitem{alv01}Alvarez, M., Shapiro, P. R., and Martel, H.,
                {\it Rev.Mex.A.A.(SC)}, in press (2001a) (astro-ph/0006203)
\bibitem{alv02}Alvarez, M., Shapiro, P. R., and Martel, H.,
                in press (this volume) (2001b)
\bibitem{exp01}Martel, H., and Shapiro, P. R., {\it Nucl.Phys.B}, {\bf 80},
                CD-Rom 09/16 (2000) (astro-ph/9904121)
\bibitem{exp02}Martel, H., and Shapiro, P. R., {\it Rev.Mex.A.A. (SC)}, 
               in press (2001a) (astro-ph/0006309)
\bibitem{mar01a}Martel, H., and Shapiro, P. R., in preparation (2001b).
\bibitem{mar01}Martel, H., Shapiro, P. R., and Valinia, A., 
               in preparation (2001)
\bibitem{nfw97}Navarro, J. F., Frenk, C. S., and White, S. D. M.,
               {\it Ap.J.}, {\bf 490}, 493 (1997)
\bibitem{val97}Valinia, A., Shapiro, P. R., Martel, H., and Vishniac, E. T. ,
                 {\it Ap.J.}, {\bf 479}, 46 (1997)
\end{references}
\end{document}